\documentstyle[aps,prl,preprint,epsfig]{revtex}
\begin{document}

\title{Exact Results for Thermodynamics of the  Classical Field Theories: Sine- and Sinh-Gordon Models}
\author{Emiliano Papa and  Alexei M. Tsvelik \\  
Department of Theoretical Physics\\ 
University of Oxford \\
1 Keble Road \\
Oxford OX1 3NP}
\maketitle

\begin{abstract}
\par
Using the recently obtained exact results for  the expectation values
of operators in the 
sine- and sinh-Gordon models 
[A. B. Zamolodchikov and S. Lukyanov, Nucl. Phys. B{\bf 493}, 571 (1997), 
V. Fateev, S. Lukyanov, A. B. Zamolodchikov and Al. B. Zamolodchikov,
Phys. Lett. B{\bf 406}, 83 (1997)] we calculate the
specific heat of the corresponding two dimensional Euclidean (classical) models. We show that the
temperature dependence of the specific heat of the sine-Gordon model, in
the commensurate phase,
has a maximum well below the Kosterlitz-Thouless transition and that the
sinh-Gordon model is  thermodynamically unstable in the strong 
coupling regime.  We give also the temperature dependence of the
specific heat in the incommensurate phase of the sine-Gordon model. 
\end{abstract}
%cond-mat/9612014


\sloppy

\section{Introduction}
 The sine-Gordon model is an exactly solvable model which has an 
enormous number of applications in
condensed matter physics and statistical mechanics. It has been
studied for years with  many remarkable results  being 
obtained. However, most of the effort has been concentrated on studying
this
model as a quantum field theory. In this paper we discuss  the
sine-Gordon model together with a less famous sinh-Gordon model as
models of classical statistical mechanics analyzing  the behavior of
their specific heat in the various range of parameters.

Let us consider  the
{\it classical} sine- and sinh-Gordon models whose partition functions are
given by
\begin{eqnarray}
%\protect
Z &=& \int {\cal{D}}\varphi e^{- S[\varphi]}\quad , \nonumber \\ 
S_{\rm sin} &\equiv& \frac{1}{T}E_{\rm sin}[\varphi]=\int d^2 x\left[
\frac{\rho_{\rm s}}{2T}\left| \nabla \varphi -
{\bf Q}\, \right|^2+ \frac{m}{T}\left(1-\cos\varphi \right)\right] \quad , \\
S_{\rm sinh} &\equiv& \frac{1}{T}E_{\rm sinh}[\varphi]=\int d^2
x\left[\frac{\rho_{\rm s}}{2T}\left| \nabla \varphi
\right|^2+ \frac{m}{T}\left(\cosh\varphi - 1\right)\right] \quad .
\end{eqnarray}
The first   model describes, for example,  the commensurate-incommensurate
transition \cite{Talapov}. Most recently it has been applied to
double-layered Quantum Hall systems \cite{MacDonald}. 

The incommensurate phase appears when $|{\bf Q}|$
exceeds some critical value and  is characterized by nonzero average
value $\left<{\bf Q}\nabla\varphi\right>$. 
Redefining the field variable $(\rho_{\rm s}/T)^{1/2}\varphi = \phi$ we reduce
the above Euclidean action to the canonical sine-Gordon form 
(see, for example, \cite{Zamolodchikov}): 
\begin{equation}
S_{\rm sin} = \int d^2x\left[\frac{1}{2}\left| \nabla \phi -
{\bf h}\beta/2\pi\, \right|^2+ \mu\left(1 - \cos\beta\phi\right)\right] \quad ,
\label{sg}
\end{equation}
with $\beta^2 = T/\rho_{\rm s}$ and $\mu = m/T$, $|{\bf h}|  = 2\pi
|{\bf Q}|/\beta^2$. 

In a similar fashion the sinh-Gordon action
becomes
\begin{equation}
S_{\rm sinh} = \int d^2x\left[\frac{1}{2}\left| \nabla \phi\, \right|^2+
\mu\left(\cosh\beta\phi - 1\right)\right] \label{sh} \quad .
\end{equation}
For this model there are no kinks and creation of nonzero field
gradient would require imaginary field $h$. We do not consider such
possibility. 

 One can consider the quantum field theory  in (1+1)-dimensions  with
$Z=\int \protect{\cal{D}} \phi e^{-S[\phi]} $ where $S$ is given by (\ref{sg}) 
or (\ref{sh}). The exact solution for both models is known and 
we can  take advantage of the fact that the free energy of 
a $D$-dimensional classical model is related to the ground state
energy $E_{\rm 0}$ of the
corresponding quantum field theory living in a space of
$(D-1)$-dimensions. More specifically, for $D = 2$ 
 the partition function of the classical theory  defined on 
a rectangle $L_x\times L_y$ with periodic boundary conditions in the 
$x$-direction at temperature $T$ is equal to the partition function of the
quantum field theory at temperature $L_x^{-1}$ with the coupling constant 
$\beta^2 = T$. The limit $L_x \rightarrow \infty$ 
corresponds to the limit of zero temperature in the quantum field theory when 
its free energy is equal to the ground state energy 
$E_{\rm 0} = L_y{\cal E_{\rm 0}}$. 
Thus we get the following relation between the free energy per unit area of the 2-dimensional classical model ${\cal F}$ 
and the ground state energy per unit length of the (1+1)-dimensional field 
theory:
\begin{equation}
{\cal F}(T) = T{\cal E}_{\rm 0}[\beta(T)] \quad .
\end{equation}
The ground state energy ${\cal E}_{\rm 0}$ of the quantum sine-Gordon model 
as a function of parameters $\beta(T)$ and $\mu$ 
is known exactly (\cite{Zamolodchikov}, \cite{Vega}) and  the corresponding
 expression for the sinh-Gordon model can be extracted from
\cite{Fateev}.

\section{Sine-Gordon Model at $Q$ = 0}

 We start our discussion of the sine-Gordon
model with the case $Q = 0$ which is already quite
nontrivial. In the Bethe ansatz approach the ground state energy of
the quantum sine-Gordon model is calculated by regularizing the model
by putting it on a lattice. The lattice constant  $a$ and the inverse
coupling constant $\theta$ of the
regularized model are  related to the mass of physical particles.
% and should be excluded from the final expression for $E_{\rm 0}$. 
According to \cite{Vega} the ground state energy  (at $Q\,=0$)  is given by:
\begin{equation}
{\cal E}_{\rm 0}=\frac{1}{a^2}\int_{-\infty}^{+\infty} \frac{\sin 4\theta t}{t}
\frac{\sinh(\pi \tau t)}{\cosh[\pi(1 - \tau) t] \ \sinh (\pi t)} dt \quad ,
\label{grounden}
\end{equation}
where $\tau =T/(8\pi \rho_{\rm s}) \equiv T/T_{\rm c}$. 
The parameters   $\theta$ and $a$ are  related to the kink's 
mass:
\begin{equation}
m_{\rm s}=\frac{4}{a} e^{- \theta/(1 - \tau)}
\quad ,
\end{equation}
To exclude $\theta$ from Eq. (\ref{grounden}) we use 
the T-dependence of  the kink's mass
given in   \cite{Zamolodchikov}:
\begin{equation}
m_{\rm s}=\frac{1}{a} \frac{2}{\sqrt{\pi}}
\frac{\Gamma{\left(\frac{1}{2}\frac{T}{T_{\rm c}-T}\right)}}{\Gamma{
\left(\frac{1}{2}\frac{T_{\rm c}}{T_{\rm c}-T}\right)}}
\left\{\frac{\pi m}{T_{\rm c}}\frac{\Gamma{\left(1-\frac{T}{T_{\rm c}}\right)}}
{\Gamma{\left(1 + \frac{T}{T_{\rm c}}\right)}}\right\}^{
\frac{1}{2\left(1-\frac{T}{T_{\rm c}}\right)}} \quad .
\end{equation}

Note that in the limiting case $\tau \ll 1$ we have 
\begin{equation}
m_{\rm s}=\frac{4}{a}\tau^{-1}(m/\pi T_{\rm c})^{1/2} \label{limit}
\quad .
\end{equation}
We should stress that the above expressions make sense 
only for $m_{\rm s}a\ll 1$ when the  continuous
approach works. Therefore to 
calculate the ground state energy  in the continuous
approximation in Eq. (\ref{grounden})
one has to keep
only the pole closest to real axis. 
% (There are two set of poles placed on
%the imaginary $t$-axis with $t=ni$ and $i(\pi/2\gamma)(1+2n)$, with 
%$n=1,2,...$ and $n=0,1,2,...$ respectively).
Near the point $\tau = 1/2$ two poles at 
$t=i$ and $t=i/2(1 - \tau)$ compete and one has to take
into account both of them. 

% Under this condition we can calculate the integral as
%sum of two parts coming from the contribution of only the nearest poles
%$t=i$ and $t=i\pi/2\gamma$. The contribution on (\ref{grounden}) from the pole
%$t=i\pi/2\gamma$ is $f_{\rm b}=(4/a^2)\cot\pi^2/2\gamma\exp{\{-2\pi
%\theta/\gamma\}}$ whereas the contribution from the
%pole $t=i$ is $f_{\rm a}=(-2/a^2)\tan\gamma \exp{\{-4\theta\}}$
%As the temperature changes the second pole
%moves in the imaginary axis from $i/2$ to $i$ and then to $+\infty$.

Taking this into account we obtain from (\ref{grounden}) 
the  general expression for the free energy:
\[
F  =  F_{\rm 1} + F_{\rm 2} \quad ,
\]
\begin{eqnarray}
\label{F1}
F_{\rm 1} & = & T\  \frac{m_{\rm s}^2}{4} \cot\left[\frac{\pi}{2 (1 - \tau)}
\right]
\\ [3mm]
& = & \frac{T}{\pi a^2}\cot\left(\frac{\pi}{2(1 - T/T_{\rm c})}\right)
\frac{\Gamma^2{\left(\frac{1}{2}\frac{T}{T_{\rm c}-T}\right)}}{\Gamma^2{
\left(\frac{1}{2}\frac{T_{\rm c}}{T_{\rm c}-T}\right)}}
\left\{\frac{\pi m}{T_{\rm c}}\frac{\Gamma{\left(1-\frac{T}{T_{\rm c}}\right)}}
{\Gamma{\left(1 +\frac{T}{T_{\rm c}}\right)}}\right\}^{
\frac{1}{\left(1-\frac{T}{T_{\rm c}}\right)}}  \quad,
\nonumber \\ [3mm]
F_{\rm 2} & = & 
 -\  T\ \frac{2}{a^{2}}\ \left(\frac{m_{\rm s}a}{4}\right)^{4(1 - \tau)}
\ \tan\left[\pi(1 - \tau)\right] \quad .
\end{eqnarray}
We emphasize that the necessity to keep both terms in the expression for the 
free energy exists only close to the free fermion point. At $\tau < 1/2$ the 
free energy remains finite in the continuous limit 
($a \rightarrow 0, m_{\rm s} =$const),  while at $\tau > 1/2$ it diverges.
The latter fact is in agreement with the perturbation theory in $m$:
\[
\int d^2x\left<\cos\beta\phi(x)\cos\beta\phi(0)\right> \sim 
\int d^2x (x/a)^{-4(1 - \tau)}
\]
diverges at small distances.
%\begin{equation}
%= - 2a^{-2}\tan[\pi(1 -\tau)]
%\left\{\frac{\Gamma\left(\frac{T}{2(T_{\rm c} -T)\right)}
%{\sqrt\pi\Gamma\left(\frac{T_{\rm c}}{2(T_{\rm c} - T)}\right)}\right\}^{4(1 -
% T/T_{\rm c})}
%\left[\frac{\pi m}{T_{\rm c}}\frac{\Gamma(1 - T/T_{\rm c})}
%{\Gamma(1 + T/T_{\rm c})}\right]^2
%\end{equation}

% The same pictures with boxes
\begin{figure}
\unitlength=1mm
\begin{picture}(165,156)
\put(-3,4){\line(1,0){162}}
\put(-3,54){\line(1,0){162}}
\put(-3,104){\line(1,0){162}}
\put(-3,4){\line(0,1){152}}
\put(159,4){\line(0,1){152}}
\put(-3,156){\line(1,0){162}}
\put(79,4){\line(0,1){152}}
\put(0,4){\epsfig{file=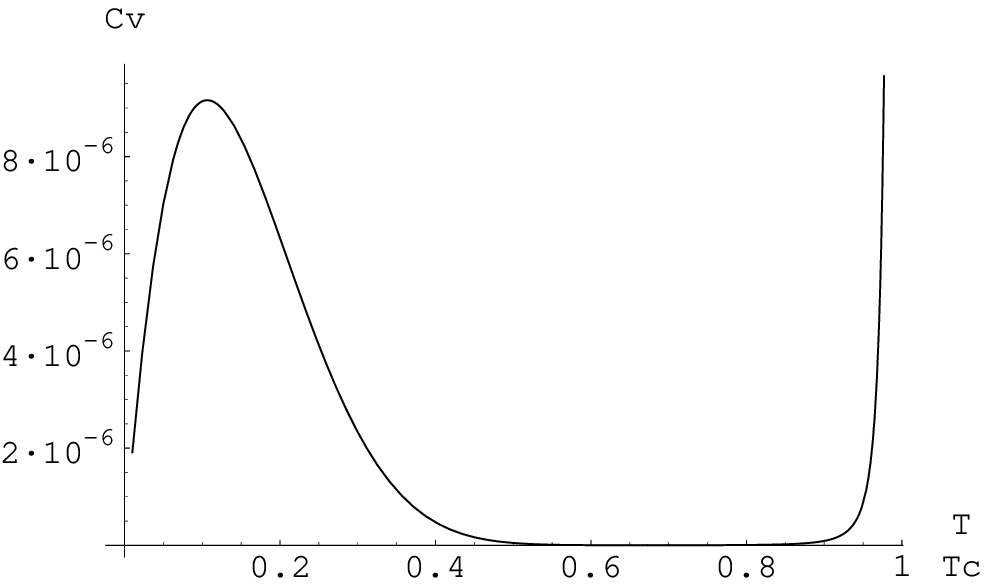,height=48mm}}
\put(0,106){\epsfig{file=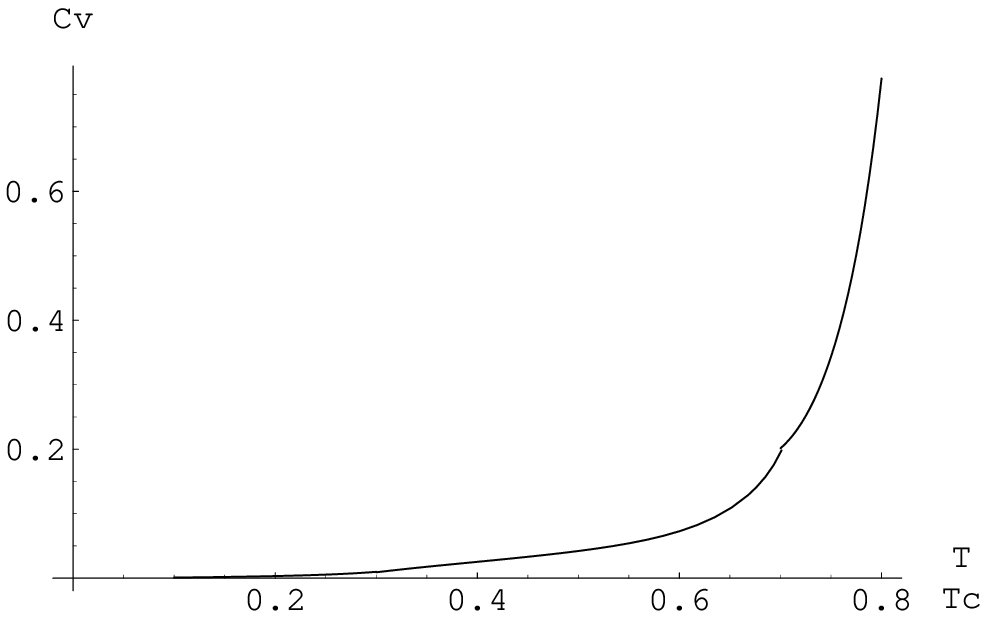,height=48mm}}
\put(81,106){\epsfig{file=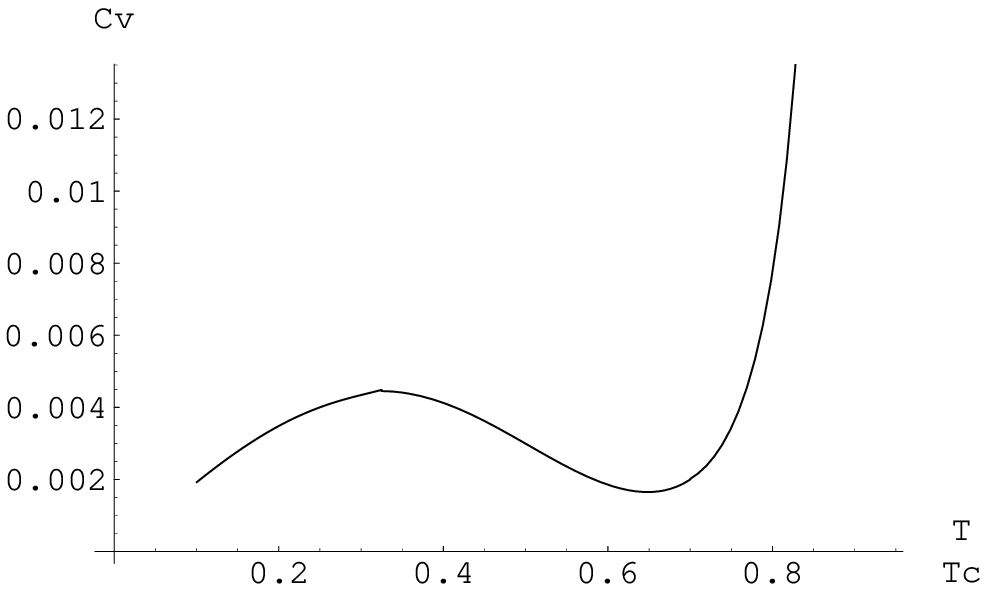,height=48mm}}
\put(0,54){\epsfig{file=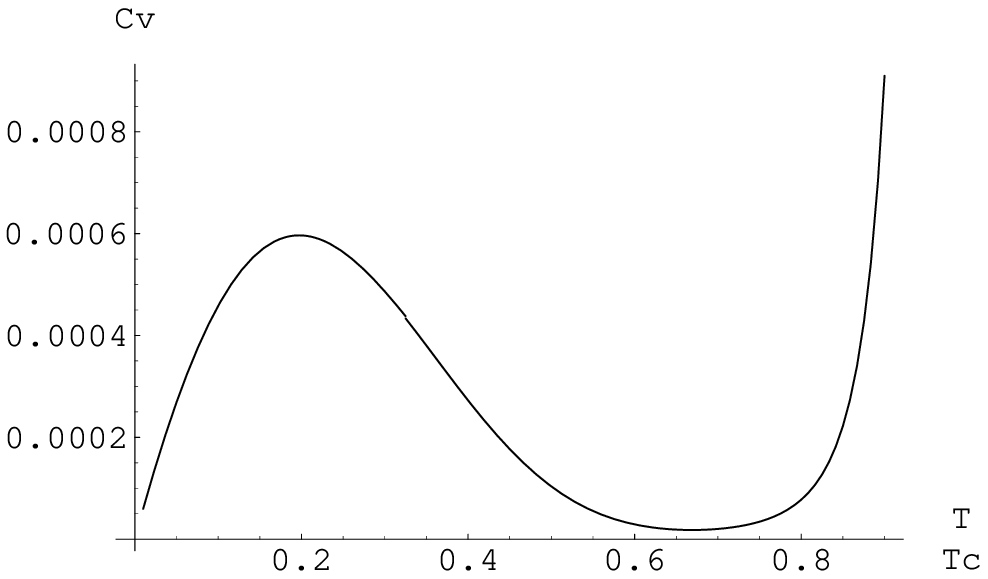,height=48mm}}
\put(81,54){\epsfig{file=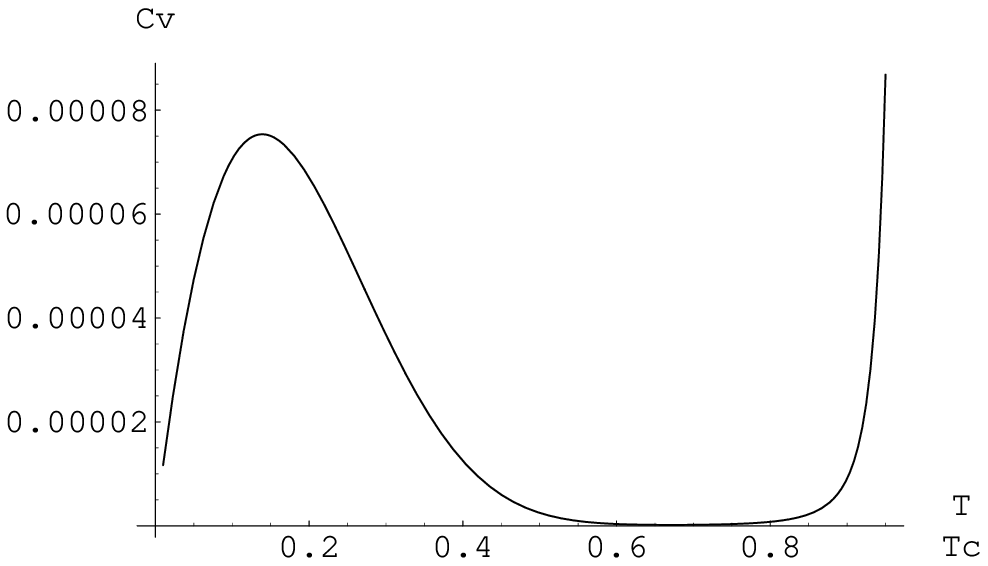,height=48mm}}
\put(81,4){\epsfig{file=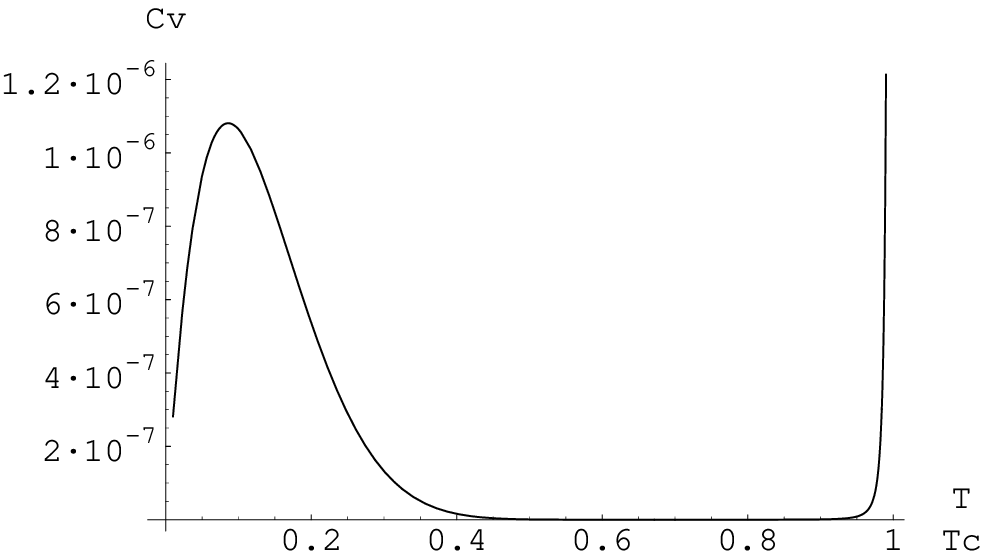,height=48mm}}
\put(40,146){\makebox(0,0)[cc]{(a)}}
\put(120,146){\makebox(0,0)[cc]{(b)}}
\put(40,95){\makebox(0,0)[cc]{(c)}}
\put(120,95){\makebox(0,0)[cc]{(d)}}
\put(40,47){\makebox(0,0)[cc]{(e)}}
\put(120,47){\makebox(0,0)[cc]{(f)}}
\put(43,140){\makebox(0,0)[tc]{$m=0.01$}}
\put(123,140){\makebox(0,0)[tc]{$m=0.001$}}
\put(47,89){\makebox(0,0)[tc]{$m=0.0001$}}
\put(127,89){\makebox(0,0)[tc]{$m=0.000\,01$}}
\put(47,41){\makebox(0,0)[tc]{$m=0.000\,001$}}
\put(127,41){\makebox(0,0)[tc]{$m=0.000\,0001$}}
\end{picture}
\caption{Plots of the specific heat as a function of $T/Tc$, for
different values of the parameter $m$, are shown.}
\end{figure}

Since we always keep $a$ finite, we plot the specific heat at finite values of
$(\pi m/T_{\rm c})$, and 
it is convenient to separate the interval of $\tau$  into 
three regions. First, for  $T/T_{\rm c}\in (0.1,0.35)$, we take the 
contribution of only the nearest pole $i/2(1 - \tau)$. 
Second, the region $T/T_{\rm c}\in(0.35,0.7)$ where we
take the contribution of both poles. Third, for 
$T/T_{\rm c}\in(0.7,0.9)$ we take only the contribution of 
the pole $t=i$.

We combine these results to find the specific heat $C_{\rm v}(T)=
-T\partial^2 F/\partial^2 T$ as a function of 
%   $T/T_{\rm c}$ and $m/T_{\rm c}$,
%   in $T/T_{\rm c}\in (0,1)$. 
\begin{figure}
\unitlength=1mm
\begin{picture}(162,56)
\put(-3,4){\line(1,0){162}}
\put(-3,56){\line(1,0){162}}
\put(-3,4){\line(0,1){52}}
\put(78,4){\line(0,1){52}}
\put(159,4){\line(0,1){52}}
\put(0,6){\epsfig{file=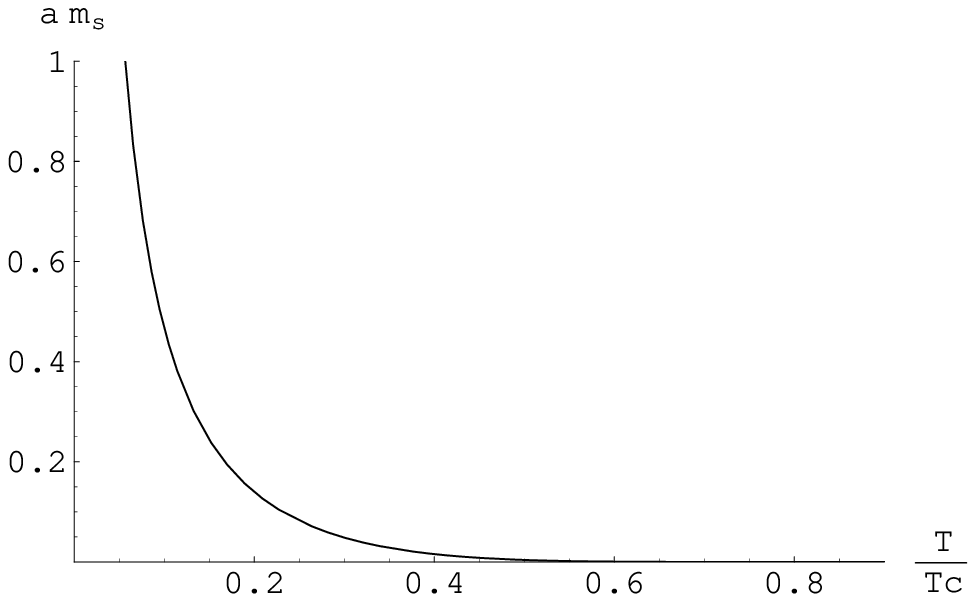,height=48mm}}
\put(81,6){\epsfig{file=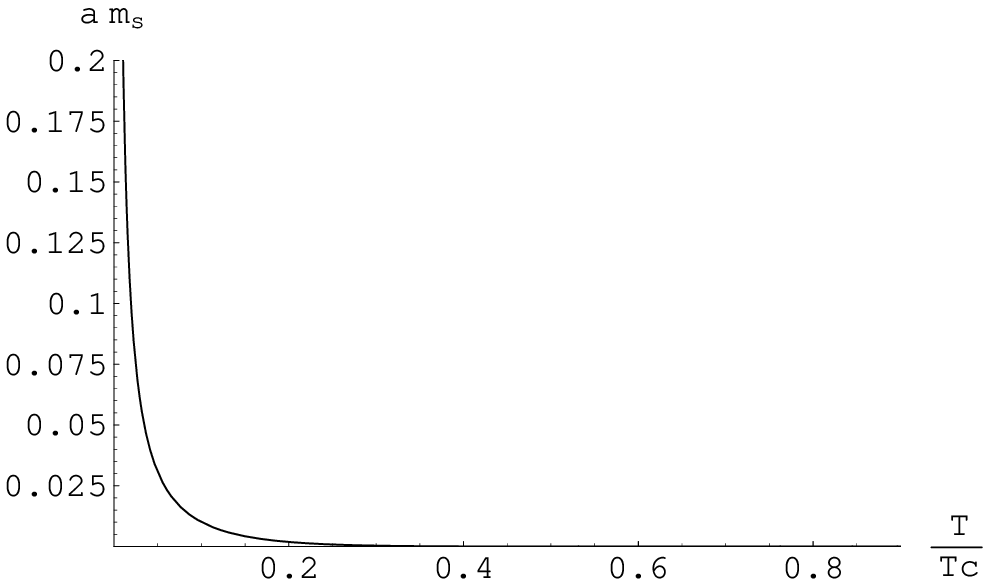,height=48mm}}
\put(47,46){\makebox(0,0)[cc]{(a)}}
\put(127,46){\makebox(0,0)[cc]{(b)}}
\put(47,40){\makebox(0,0)[tc]{$m=0.0001$}}
\put(127,40){\makebox(0,0)[tc]{$m=0.000 0001$}}
\end{picture}
\caption{Plots of $(a m_{\rm s})$ as function of $(T/Tc)$, for different
values of the parameter $m$, are shown. The condition of continuous
approach works on the region $(a m_{\rm s})\ll 1$.}
\end{figure}
  $T/T_{\rm c}$ and $m/T_{\rm c}$,
  in $T/T_{\rm c}\in (0,1)$.
On Fig. 1  we present  plots of the temperature of the specific heat
for various values of $m$. Fig. 2 gives temperature dependence 
of  the kink's mass. From this picture one one can estimate the region where
the condition $m_{\rm s}a\ll 1$ is fulfilled. 

At $1 - \tau \ll 1$ and at $\tau \ll 1$  expressions (\ref{F1}) for the
specific heat simplify. In the first case we have 
\begin{equation}
a^2 F = - \frac{2\pi^3 m^2}{T_{\rm c}}\left\{\frac{1}{1 - \tau} +
2\ln\left[\frac{e^2}{\pi(1 - \tau)}\right] + O\left((1 - \tau)\ln^2(1 -
\tau)\right)\right\} \quad ,
\end{equation}
\begin{equation}
C_{\rm v} = \frac{4\pi^3}{a^{2}}\frac{m^2T_{\rm c}}{(T_{\rm c} - T)^3} + ...
\quad .
\end{equation}
This singularity is associated with the Kosterlitz-Thouless transition
at $T = T_{\rm c}$.

At $\tau \ll 1$ we have  
\begin{eqnarray} 
a^2 F &=& - 4\pi^2 m\left(\frac{\pi m}{T_{\rm c}}\right)^{1/(1 - \tau)}
\quad , \\
C_{\rm v} &=& 4\pi\left(\frac{\pi m}{T_{\rm c}}\right)^2
\ln\left(\frac{T_{\rm c}}{\pi m}\right)\left[2 +
\ln\left(\frac{T_{\rm c}}{\pi m} \right)\right] \tau
\exp\left\{- \tau\ln\left(\frac{T_{\rm c}}{\pi m} \right)\right\} \quad .
\end{eqnarray}
The latter expression explains the existence of the maximum in the
specific heat: at $\ln(T_{\rm c}/\pi m) \gg 1$ the maximum occurs at
$\tau^* = [\ln(T_{\rm c}/\pi m)]^{-1}$.

\section{Sine-Gordon Model in the Incommensurate Phase}

At 
\begin{equation}
Q > Q_{\rm c} = 4\tau m_{\rm s}(\tau) \quad \label{condit}
\label{criteq}
\end{equation}
the sine-Gordon model is in the 
incommensurate phase characterized by a condensate of 
kinks $\left<{\bf Q}\nabla\phi \right> \neq 0$. The ground state
energy of the corresponding quantum field theory acquires an additional
contribution originating from the condensate. The corresponding change
 in the free energy of the classical model is 
\begin{eqnarray}
\delta F = \frac{\rho_{\rm c} Q^2}{2} + \frac{T m_{\rm s}}{2\pi}\int_{-B}^B
d\theta\cosh\theta \epsilon(\theta) \quad .
\end{eqnarray}
The nonpositive function $\epsilon(\theta)$ is defined inside the 
interval $-B< \theta < B$ and satisfies the integral equation (see,
for example \cite{Zamolodchikov})
\begin{eqnarray}
\epsilon(\theta) + \int_{-B}^B d\theta' K(\theta -
\theta')\epsilon(\theta') = m_{\rm s}\cosh\theta - \frac{Q}{4\tau} \quad ,
\label{integ}
\end{eqnarray}
where the Fourier image of the kernel is
\[
K(\omega) = \frac{\sinh\frac{\pi(1 - 2\tau)\omega}{2(1 -
\tau)}}{2\cosh(\pi\omega/2)\sinh\frac{\pi\tau\omega}{2(1 - \tau)}} \quad .
\] 
The kernel $K(\theta)$ encodes the information about the soliton-soliton
scattering.
The limit $B$ is determined  by the condition $\epsilon(\pm B) = 0$. A
possibility of $\epsilon$ being negative appears when the right hand
side becomes not positively defined which corresponds to condition
(\ref{condit}). 
  
We have solved the integral equation for the function $\epsilon(\theta)$ 
numerically and the plots of its dependence on $\tau$ and $Q$ are shown in
Fig. \ref{epstheta}.
 The dependence of
the limit $B$ on the temperature and $Q$ is shown on  
Fig. \ref{boundaryB}.

\begin{figure}
\unitlength=1mm
\begin{picture}(160,55)
%\put(0,-3){\epsfig{file=critline1.eps,height=50mm}}
\put(-3,-3){\epsfig{file=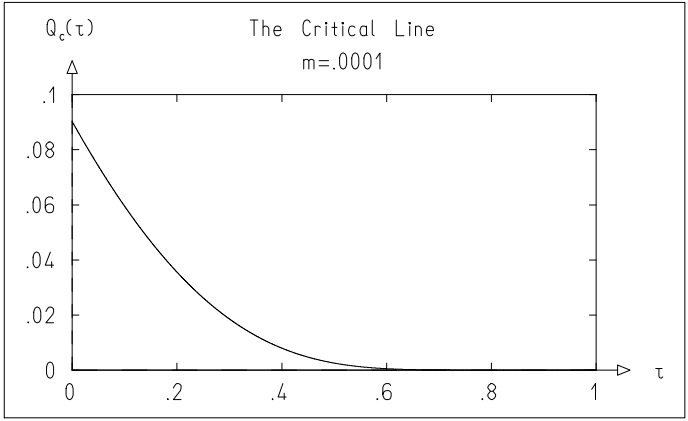,height=55mm}}
\put(92,0){\makebox(68,55)[b]{\begin{minipage}[b]{68mm}
\protect\caption{\sloppy  % kjo eshte qe fjalet te ndahen mire ne caption
The critical line as function of $(T/T_{\rm c})$  for a
fixed value of the parameter $m$ is shown. The lattice constant $a$ is
taken to be unity.}
%\label{figGaus}}
\end{minipage}}}
\end{picture}
\end{figure}

It is curious that the critical field has a finite limit at $\tau
\rightarrow 0$ (this limit was first considered in \cite{fowler}). According to Eq. (\ref{limit}), we have 
\begin{equation}
aQ_{\rm c}(0) = 16\left(\frac{m}{\pi T_{\rm c}}\right)^{1/2} \quad .
\end{equation}
In the vicinity of the critical line in the incommensurate phase 
one can expand solution of
Eq. (\ref{integ}) in series in $B$ and get for the additional free energy:
\begin{eqnarray}
\frac{1}{T_{\rm c}}\delta F & = & \frac{Q^2}{2} 
\\ [3mm]
& & - \frac{\tau m_{\rm s}^2}{6\pi}
\left(\frac{Q}{Q_{\rm c}}- 1\right)^{3/2}\left[1 -
K(0)\left(\frac{Q}{Q_{\rm c}}- 1\right)^{1/2} + \left(0.1 +
7 \frac{K^2(0)}{6}\right)\left(\frac{Q}{Q_{\rm c}}- 1\right) + ...\right]\ .
\nonumber
\end{eqnarray}
At small $\tau$, $K(0) \approx \frac{1}{\pi^2\tau}\ln(1/\tau)$ and the
expansion is valid for 
\begin{equation}
\pi^2\tau/\ln(1/\tau) \gg (Q/Q_{\rm c} - 1) \label{cond} \quad .
\end{equation}
Plots of the additional specific heat are shown on Fig. \ref{Cvalltogether}a),
whereas plots of the total specific heat are shown in 
Fig. \ref{Cvalltogether}b) for some values
of the field $Q$.

The  $Q/\tau \rightarrow \infty$ analytic structure of the total free
energy $F(Q)$ is given by Zamolodchikov \cite{Zamolodchikov}
\begin{equation}
\label{eq22}
F(Q)-F(0) = \frac{\rho_{\rm c} Q^2}{2} - T\  \frac{m_{\rm s}^2}{4}
\cot\left[\frac{\pi}{2 (1 - \tau)} \right] 
- \left(\frac{Q}{4 \tau}\right)^2 \frac{k(Q/4\tau)}{\pi} \quad .
\end{equation}
The factor $k(Q/4\tau)$ is given as a power series
\begin{figure}
\unitlength=1mm
\begin{picture}(160,103)
\put(-2,53){\epsfig{file=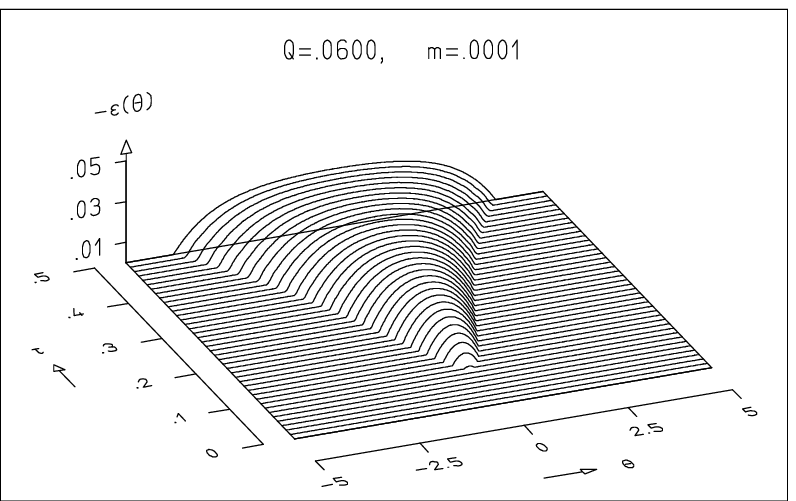,height=50mm}}
\put(78,53){\epsfig{file=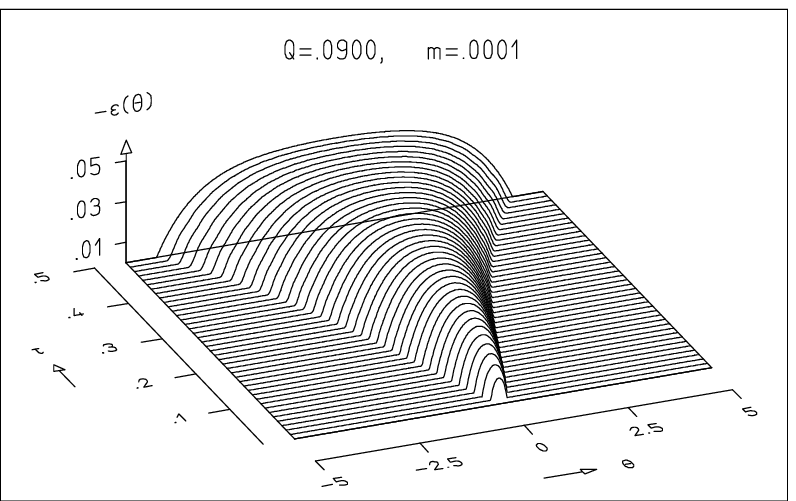,height=50mm}}
%\put(78,53){\epsfig{file=critline1.eps,height=50mm}}
\put(-2,3){\epsfig{file=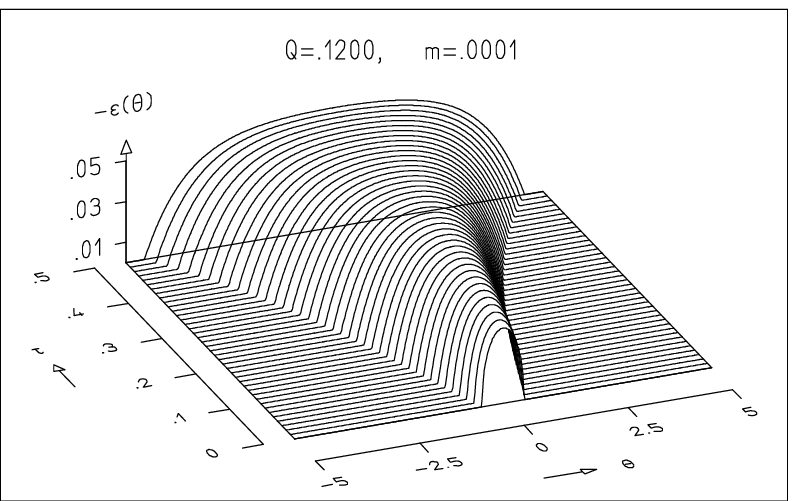,height=50mm}}
\put(78,3){\epsfig{file=IQbaa.MET,height=50mm}}
\put(6,99){\makebox(0,0)[cc]{(a)}}
\put(150,99){\makebox(0,0)[cc]{(b)}}
\put(6,49){\makebox(0,0)[cc]{(c)}}
\put(150,49){\makebox(0,0)[cc]{(d)}}
\end{picture}
\caption{(a) Plots of $(-\epsilon(\theta))$ for different fixed
$\tau$. On  figure (b) the parameter $Q = 0.090$,  which corresponds to
crossing
the critical line at $\tau\approx 0$.}
\label{epstheta}
\end{figure}
\begin{figure}
\unitlength=1mm
\begin{picture}(160,53)
\put(-2,3){\epsfig{file=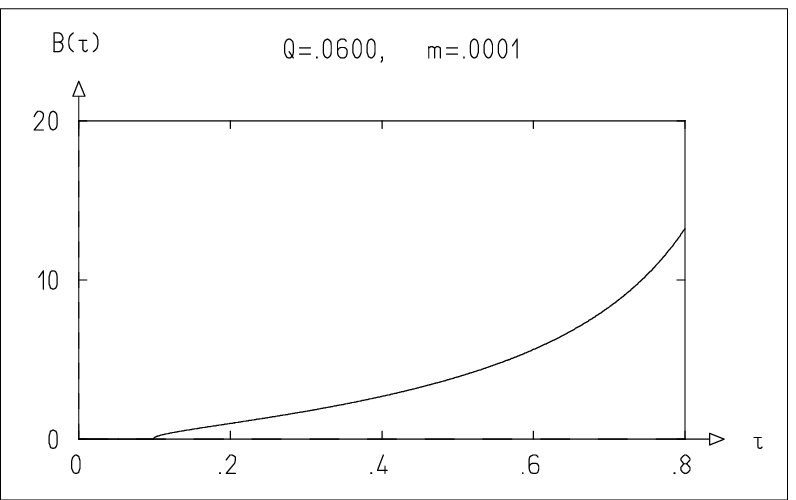,height=50mm}}
\put(78,3){\epsfig{file=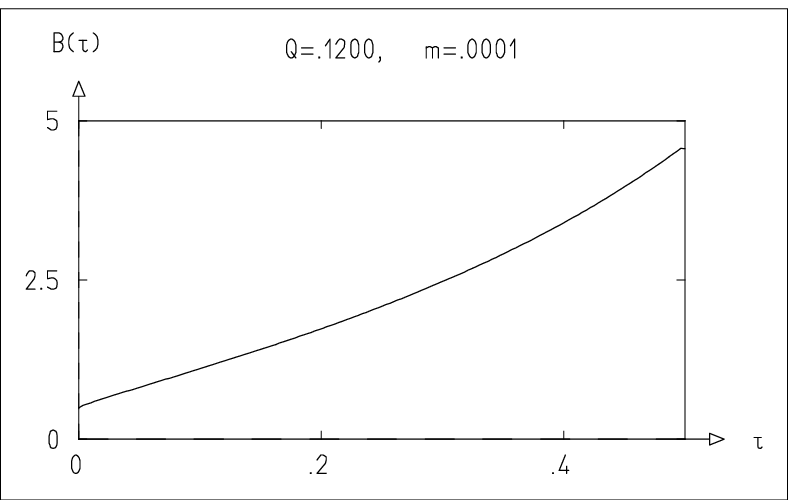,height=50mm}}
%\put(6,49){\makebox(0,0)[cc]{(a)}}
%\put(87,49){\makebox(0,0)[cc]{(b)}}
\end{picture}
\caption{Plots of temperature dependence of the parameter  $B$ in
Eq. (\ref{integ}).
The parameter $Q$ is taken to have values $0.060$ and $0.120$,
respectively.}
\label{boundaryB}
\end{figure}
\begin{equation}
k(Q/4\tau)=\sum_{n=0}^{\infty}K_{\rm n}y^n \quad ,
\end{equation}
with 
\begin{equation}
y = \left[\frac{2 m_{\rm s}\sqrt{\pi}}{Q}\frac{\Gamma\left(\frac{1}{2(1-\tau)}
\right)}{\Gamma\left(\frac{\tau}{2(1-\tau)}\right)}\right]^{4(1-\tau)}
\quad .
\end{equation}

\begin{figure}
\unitlength=1mm
\begin{picture}(160,53)
\put(77,3.1){\line(1,0){82}}
\put(159,3.1){\line(0,1){50}}
\put(77,53.1){\line(1,0){82}}
%\put(-3,3){\epsfig{file=IQEMI.MET,height=50mm}}
\put(-3,3){\epsfig{file=IQLAS.MET,height=50mm}}
\put(78,3){\epsfig{file=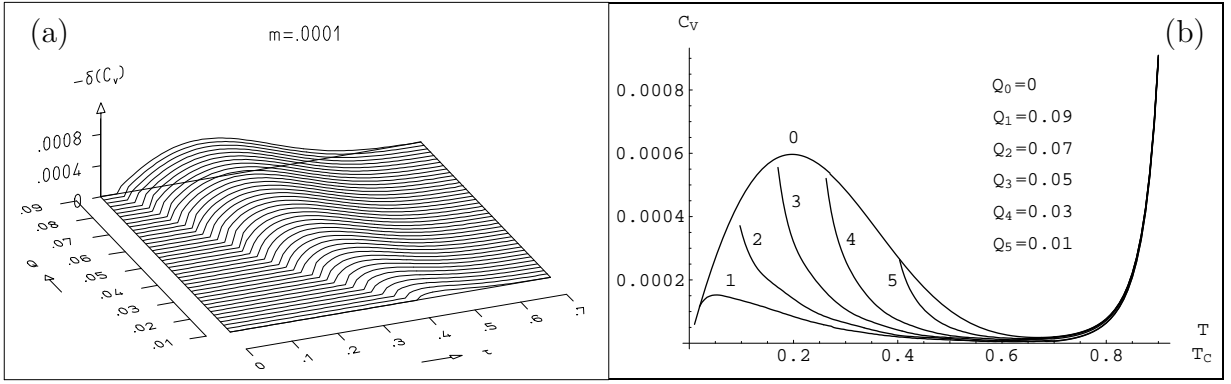,height=50mm}}
%\put(78,3){\epsfig{file=Cvall.eps,height=50mm}}
\put(3,49){\makebox(0,0)[cc]{(a)}}
%\put(6,43){\makebox(0,0)[cc]{-}}
\put(154,49){\makebox(0,0)[cc]{(b)}}
\end{picture}
\caption{(a) Three dimensional plot of the contribution to the
specific heat coming from the soliton
condensate
%$-\delta(C_{\rm v})$
as function of
$\tau$ and $Q$. The presence of the condensate
leads to decrease in the specific heat. (b) The total specific heat of
the
system for different values of the parameter $Q$.
Each of the lines 1-5 merges with  the line 0 ($Q=0$) at the
commensurate-incommensurate  transition temperature.}
\label{Cvalltogether}
\end{figure}

The first two coefficients are given by
\begin{eqnarray}
%\lefteqnarray{
K_{\rm 0}  =  \tau \quad ,    \qquad 
K_{\rm 1}  =  2 \frac{\Gamma(\tau)}{\Gamma(-\tau)} 
\frac{\Gamma(5/2-\tau)}{\Gamma(1/2+\tau)}
\frac{\tau}{(2\tau-1)(3-2\tau)} \quad .
\end{eqnarray}
Putting them together in Eq. (\ref{eq22}) we get
\begin{eqnarray}
F(Q) -F(0) & = & \frac{\rho_{\rm c} Q^2}{2}- T\  \frac{m_{\rm s}^2}{4}
\cot\left[\frac{\pi}{2 (1 - \tau)} \right]
-T\left(\frac{Q}{4 \tau}\right)^2 \frac{\tau}{\pi}
\\ [3mm]
& & -T\ \frac{2}{\pi}\left(\frac{Q}{4 \tau}\right)^2
\frac{\Gamma(\tau)}{\Gamma(-\tau)}\frac{\Gamma(5/2-\tau)}{\Gamma(1/2+\tau)}
\frac{\tau}{(2\tau-1)(3-2\tau)}\left[\frac{2 m_{\rm
s}\sqrt{\pi}}{Q}\frac{\Gamma\left(\frac{1}{2(1-\tau)}
\right)}{\Gamma\left(\frac{\tau}{2(1-\tau)}\right)}\right]^{4(1-\tau)} \ .
\nonumber 
\end{eqnarray}
The second term on the right hand side of the above equation gives the
free energy of the system in absence of the field ${\bf Q}$ (with
opposite sign) and  cancels with $F(0)$ of the left hand side. 
On the other hand the first and the third terms on the right 
cancel each other (keeping in mind that $T_{\rm c}=8 \pi \rho_{\rm c}$).
The total free energy of the system in the presence of the field ${\bf
Q}$ is given, to this order of approximation, by:
\begin{eqnarray}
F(Q) & = & -T\ \frac{2}{\pi}\left(\frac{Q}{4 \tau}\right)^2
\frac{\Gamma(\tau)}{\Gamma(-\tau)}\frac{\Gamma(5/2-\tau)}{\Gamma(1/2+\tau)}
\frac{\tau}{(2\tau-1)(3-2\tau)}\left[\frac{2 m_{\rm s}
\sqrt{\pi}}{Q}\frac{\Gamma\left(\frac{1}{2(1-\tau)}
\right)}{\Gamma\left(\frac{\tau}{2(1-\tau)}\right)}\right]^{4(1-\tau)} 
\nonumber \\ [3mm] 
& = & 
-Q^{-2+4\tau}\ \frac{T_{\rm c}}{8 \pi} 
\frac{\Gamma(\tau)}{\Gamma(-\tau)}\frac{\Gamma(5/2-\tau)}{\Gamma(1/2+\tau)}
\frac{1}{(2\tau-1)(3-2\tau)}\left[2 m_{\rm s} \sqrt{\pi} \
\frac{\Gamma\left(\frac{1}{2(1-\tau)}
\right)}{\Gamma\left(\frac{\tau}{2(1-\tau)}\right)}\right]^{4(1-\tau)}
\ .
\end{eqnarray}
The expansion is valid only for small values of $\tau < 1/2$, where the
free energy of the system in absence of the field $Q$ is given by
$F_{\rm 1}$ of Eq. (\ref{F1}).

%Taking into account Eq. (\ref{limit}) we get at $\tau \ll 1$ with
%condition (\ref{cond}) fulfilled
%\begin{equation}
%\delta C_{\rm v} \approx \frac{m T_{\rm c}}{3T^2}\left(\frac{Q}{Q_{\rm c}}-
%1\right)^{3/2} \label{cv}
%\end{equation}
%At $\tau \rightarrow 0$ the kernel $K(\theta)$ becomes singular at
%$\theta \rightarrow 0$ and Eq. (\ref{integ}) becomes 
%\begin{equation}
%\frac{1}{\pi^2}\int_{-B}^B d\theta' \ln|\coth[(\theta  -
%\theta')/2]|\epsilon(\theta') = 2(\pi m T_{\rm c})^{1/2}\left(\cosh\theta -
%\frac{Q}{Q_{\rm c}(0)}\right)
%\end{equation}
%with 
%\[
%\tilde K(\omega) = \frac{\tanh(\pi\omega/2)}{\pi\omega}
%\]
%and 
%\begin{equation}
%\delta F = \frac{\rho_{\rm c} Q^2}{2} + 
%\frac{(\pi T_{\rm c} m)^{1/2}}{\pi}\int_{-B}^B
%d\theta\cosh\theta \epsilon(\theta)
%\end{equation}
%Thus we see that temperature disappears from the equations. Thus we
%conclude that the corresponding contribution to the specific heat
%vanishes at $T \rightarrow 0$. Therefore we conclude that the $\delta
%C_{\rm v}$ has maximum when condition (\ref{cond}) is about to break down. 
%Assuming that Eq. (\ref{cv})  still
%gives the correct order of magnitude we get 
%\begin{equation}
%\max(\delta C_{\rm v}) \approx \frac{\pi^3 m}{3T_{\rm c}}(Q/Q_{\rm c} -
%1)^{1/2}   
%\end{equation}

\section{Sinh-Gordon Model; Thermodynamic Instability}

 Now we  consider the sinh-Gordon model. Here the exact 
solution was suggested by Fateev  \cite{Fateev} who has taken the
sine-Gordon 
two-body S-matrix for the first breathers and changed the sign of the
coupling constant $\beta^2$ in it. Comparing Eqs.(\ref{sg}, \ref{sh}) we 
see that to get the sinh-Gordon action out of the sine-Gordon one we have to 
change  $\beta \rightarrow i\beta$ and $m \rightarrow - m$. 

 Doing this substitution in the expression for
 the free energy (see also 
 \cite{Fateev}), we get 
\begin{eqnarray}
F = - mI(T/T_{\rm c})(1 + T/T_{\rm c})\left[\frac{T_{\rm c} 
\Gamma(1 - T/T_{\rm c})}{\pi
m\Gamma(1 + T/T_{\rm c})}\right]^{\frac{T}{T + T_{\rm c}}} \label{Fsinh}
\quad ,
\end{eqnarray}
\begin{eqnarray}
I(x) = \exp\left\{2\int_0^{\infty} \frac{dt}{t}\left[- \frac{\cosh^2(xt)\sinh(xt)}{\sinh
t\cosh((1 + x)t)} + xe^{-2t}\right]\right\} \label{I}
\quad ,
\end{eqnarray}
where $T_{\rm c} = 8\pi\rho_{\rm s}$. This expression is valid for
$T < T_{\rm c}$. 
The integral (\ref{I}) can be calculated to give the expression
\begin{equation}
I(x)=\frac{\Gamma\left(\frac{1}{2}+\frac{x}{2(1+x)}\right)}
{\Gamma\left(\frac{1}{2}-\frac{x}{2(1+x)}\right)}
\frac{\Gamma(x)}{\Gamma(-x)}
\frac{\Gamma\left(-\frac{x}{2(1+x)}\right)}
{\Gamma\left(\frac{x}{2(1+x)}\right)}
\quad ,
\end{equation}
with $0<x<1$. $I(x)$ is monotonically decreasing function taking values in
the interval $(0,1)$.

In the light of the following discussion it will be instructive also
to have the expression for the mass (the inverse correlation length)
of the sinh-Gordon theory. To get this expression one has to change
$\beta$ to $i\beta$ for the sine-Gordon mass which corresponds to 
reversal of  the sign of  $T_{\rm c}$ in Eq.(8):
\begin{eqnarray}
m_{\rm s} a = \frac{4\sqrt\pi}{\Gamma\left(\frac{T_{\rm c}}{2(T + T_{\rm c})}
\right)\Gamma\left(1 +
\frac{T}{2(T + T_{\rm c})}\right)}\left[\frac{\pi m}{T_{\rm c}}
\frac{\Gamma\left(1 + T/T_{\rm c}\right)}{\Gamma\left(1 - T/T_{\rm c}\right)}
\right]^{\frac{T_{\rm c}}{2(T + T_{\rm c})}} \label{masssinh}
\quad .
\end{eqnarray}
(This expression coincides with the mass of the sine-Gordon {\it breather}
after the substitution $T \rightarrow -T$). 
 We would like to attract the reader's attention to the fact that $m(T)$
never diverges at $T < T_{\rm c}$ and actually goes to zero at $T
\rightarrow T_{\rm c}$. 
 
 For $\tau = (1 - T/T_{\rm c}) \ll 1$ we obtain from 
(\ref{I}) $I(\tau) \sim \tau$ and substituting this into
Eq. (\ref{Fsinh}) we get at $\ln(T_{\rm c}/\pi m) \gg 1$ 
\begin{equation}
F \sim - (1 - T/T_{\rm c})^{1/2}e^{- (1 - T/T_{\rm c})\ln(T_{\rm c}/\pi m)}
\quad .
\end{equation}
It is easy to see that the specific heat  becomes negative  at 
\begin{equation}
(1 - T/T_{\rm c}) < (1 + \sqrt 2) [2\ln(T_{\rm c}/\pi m)]^{-1} \label{inst}
\quad ,
\end{equation}
such that and at  temperatures sufficiently close 
 to $T_{\rm c}$ the model is thermodynamically unstable. 
%Near $T_{\rm c}$ we have 
%\begin{equation}
% C_{\rm v} \sim (1 - T/T_{\rm c})^{-3/2}
%\end{equation} 
%which together with disappearance of the mass 
%signifies a second order phase transition. 

 This  thermodynamic instability is not completely unexpected. 
It occurs  in the strong coupling regime when  the coupling 
constant $T/T_{\rm c}$ approaches its critical value 1. Let the reader recall
that at this value of the coupling constant 
the ultraviolet limit of the sinh-Gordon model - the Liouville model becomes 
unstable \cite{zamol2} (its central charge approaches the value of 25). 
Another indication of the instability comes from  
the temperature dependence of the inverse correlation length given by
Eq. (\ref{masssinh}): at $T \rightarrow T_{\rm c}$ the mass becomes zero which
one does not expect to happen to the sinh-Gordon model which can be thought as
the Gaussian theory perturbed by a strongly relevant operator. One 
can check that at the 
instability point (\ref{inst}) the mass of the sinh-Gordon particle 
is still much larger than its value at $ T = 0$.  

\section{Conclusions}

The summary of our results on the classical sine-Gordon model is well
represented by Fig. 6. The specific heat has a peak well below the
Kosterlitz-Thouless transition and the temperature dependence becomes
even more complicated in the incommensurate phase. We suppose that 
all these features are detectable  experimentally in the relevant
systems like the one described in \cite{MacDonald}.

\newpage

{\bf Acknowledgments}

E. P. would like to thank Prof. H. D. Dahmen for the hospitality he offered
during the stay at the University of Siegen, Germany, where part of this work
was done. He also would like to thank Dave Allen, Joseph Betouras, 
Pedro Ferreira and especially Tilo Stroh for many helpful discussions. 
%%and to acknowledge the financial support of the Bodossaki and Dulverton 
%%foundations and the ORS Award Scheme.
A. M. T. is grateful to H. Saleur, F. Smirnov  V. Fateev 
and above all to Yu Lu  
for stimulating discussions and valuable remarks.

\end{document}